
\input harvmac
%
%
%
%
\ifx\answ\bigans
\else
\output={
  \almostshipout{\leftline{\vbox{\pagebody\makefootline}}}\advancepageno
}
\fi
%
%
%
\def\mayer{\vbox{\sl\centerline{Department of Physics 0319}%
\centerline{University of California, San Diego}
\centerline{9500 Gilman Drive}
\centerline{La Jolla, CA 92093-0319}}}
%
%

%
%
\def\UCSD#1#2{\noindent#1\hfill #2%
\bigskip\supereject\global\hsize=\hsbody%
\footline={\hss\tenrm\folio\hss}}
%
%
\def\abstract#1{\centerline{\bf Abstract}\nobreak\medskip\nobreak\par #1}
%
%
%
%
\edef\tfontsize{ scaled\magstep3}
 \tfontsize  \tfontsize
 \tfontsize \font\titlei=cmmi10 \tfontsize
\font\titleis=cmmi7 \tfontsize \font\titleiss=cmmi5 \tfontsize
\font\titlesy=cmsy10 \tfontsize \font\titlesys=cmsy7 \tfontsize
\font\titlesyss=cmsy5 \tfontsize  \tfontsize
\skewchar\titlei='177 \skewchar\titleis='177 \skewchar\titleiss='177
\skewchar\titlesy='60 \skewchar\titlesys='60 \skewchar\titlesyss='60
%
%
%
%
%
\def\inv{^{\raise.15ex\hbox{${\scriptscriptstyle -}$}\kern-.05em 1}}
\def\pr#1{#1^\prime}  
\def\lbar{{\lower.35ex\hbox{$\mathchar'26$}\mkern-10mu\lambda}} 

%
%
%
%
\def\dsl{\,\raise.15ex\hbox{/}\mkern-13.5mu D} 
\def\delsl{\raise.15ex\hbox{/}\kern-.57em\partial}
\def\Ksl{\hbox{/\kern-.6000em\rm K}}
\def\Asl{\hbox{/\kern-.6500em \rm A}}
\def\Dsl{\hbox{/\kern-.6000em\rm D}} 
\def\Qsl{\hbox{/\kern-.6000em\rm Q}}
\def\gradsl{\hbox{/\kern-.6500em$\nabla$}}
%
%
\def\lspace{\ifx\answ\bigans{}\else\qquad\fi}
\def\lbspace{\ifx\answ\bigans{}\else\hskip-.2in\fi} 
%
%
\def\boxeqn#1{\vcenter{\vbox{\hrule\hbox{\vrule\kern3pt\vbox{\kern3pt
        \hbox{${\displaystyle #1}$}\kern3pt}\kern3pt\vrule}\hrule}}}
%
%
\def\mbox#1#2{\vcenter{\hrule \hbox{\vrule height#2in
\kern#1in \vrule} \hrule}}
%
%
%
%

   \def\CL{{\cal L}}
  \def\CO{{\cal O}}

%
%
%
%
%

%

\def\bar#1{\overline{#1}}
\def\vev#1{\left\langle #1 \right\rangle}

\def\ket#1{\left| #1\right\rangle}

\def\darr#1{\raise1.5ex\hbox{$\leftrightarrow$}\mkern-16.5mu #1}

%
%
\def\frac#1#2{{\textstyle{#1\over #2}}} 
%
%
%
%

%
%
%
%

%
%
\def\ltap{\ \raise.3ex\hbox{$<$\kern-.75em\lower1ex\hbox{$\sim$}}\ }
\def\gtap{\ \raise.3ex\hbox{$>$\kern-.75em\lower1ex\hbox{$\sim$}}\ }
\def\gl{\ \raise.5ex\hbox{$>$}\kern-.8em\lower.5ex\hbox{$<$}\ }
\def\roughly#1{\raise.3ex\hbox{$#1$\kern-.75em\lower1ex\hbox{$\sim$}}}
%
%
\def\ie{\hbox{\it i.e.}}        
        
\def\etal{\hbox{\it et al.}}

\def\np#1#2#3{{Nucl. Phys. } B{#1} (#2) #3}
\def\pl#1#2#3{{Phys. Lett. } {#1}B (#2) #3}
\def\prl#1#2#3{{Phys. Rev. Lett. } {#1} (#2) #3}
\def\physrev#1#2#3{{Phys. Rev. } {#1} (#2) #3}

\def\cmp#1#2#3{{Comm. Math. Phys. } {#1} (#2) #3}
\relax

\def\CO{{\cal O}}

\def\lta{\ \hbox{\raise.55ex\hbox{$<$}} \!\!\!\!\!
\hbox{\raise-.5ex\hbox{$\sim$}}\ }
\def\gta{\ \hbox{\raise.55ex\hbox{$>$}} \!\!\!\!\!
\hbox{\raise-.5ex\hbox{$\sim$}}\ }
\def\mayer{\vbox{\sl\centerline{Department of Physics}
\centerline{9500 Gilman Drive 0319}
\centerline{University of California, San Diego}
\centerline{La Jolla, CA 92093-0319}}}

\def\noblackbox{\overfullrule=0pt}
\def\ket#1{\vert #1\rangle}
\hyphenation{anom-aly anom-alies coun-ter-term coun-ter-terms}
\relax
\def\lorentz{\matrix{
1&&&\cr
&-1&&\cr
&&-1&\cr
&&&-1\cr}}
\def\mzer{\matrix{\ &\ &\ \cr&0&\cr&&\cr}}
\def\mK{\matrix{&&\cr&K_L&\cr&&\cr}}
\def\cp{$\widehat{{\rm CP}}$ }
\noblackbox
 \centerline{{\titlefont{ Is CP a  Gauge Symmetry? }}}
\bigskip
\centerline{\it Submitted to Nuclear Physics B}
\vskip .5in
\centerline{Kiwoon Choi,
David B. Kaplan\footnote{$^a$}{NSF Presidential
Young Investigator, DOE Outstanding Junior
Investigator}\footnote{$^b$}{Sloan Fellow} and  Ann E. Nelson
$^b$\footnote{$ ^c$}{SSC Fellow}\footnote{\ }{E-mail:
kchoi@ucsd.bitnet,\ dkaplan@ucsd.bitnet,\
anelson@ucsd.bitnet}} \bigskip \medskip
\mayer
\vfill
\abstract{
Conventional solutions to the strong CP problem all require the
existence of   global symmetries.  However quantum
gravity may destroy global symmetries, making it hard to understand
why the electric dipole moment of the neutron (EDMN) is so small.
We suggest here that CP is actually a discrete {\it gauge}
symmetry, and is therefore not violated by quantum gravity.
We show that four dimensional CP can arise as a discrete gauge
symmetry in
theories with dimensional compactification, if the original
number of Minkowski dimensions   equals $8k+1$, $8k+2$ or $8k+3$,
and if there are  certain restrictions on the gauge group; these
conditions are
met by superstrings.
CP may  then be  broken spontaneously below $10^9$
GeV, explaining the observed CP violation in the kaon system without
inducing a large  EDMN.   We discuss the phenomenology of such
models, as well as the peculiar properties of  cosmic ``CP
strings'' which could be produced at the compactification scale.
Such strings have the curious property that a particle carried
around the string is turned into its CP conjugate. A single  CP
string renders  four dimensional spacetime nonorientable.
  }
\vfill
\UCSD{UCSD/PTH 92-11}{April 1992}

\newsec{Motivation: the strong CP problem}

We still do not know why CP   is violated by the weak interactions
but not by the strong.  In order to satisfy the upper bound  of
$1.2\times 10^{-25} $ e-cm on the neutron electric dipole moment
\ref\edmn{ K.F. Smith \etal, \pl{234}{1990}{191}  } the strong CP
violating parameter  $\bar\theta$  \eqn\thetadef{\bar\theta\equiv
\theta_{\rm QCD}+{\rm arg}\;{\rm det}\; m_q} must satisfy
$\bar\theta\ltap 2 \times 10^{-10} $ for  the usually quoted up to
down quark ratio  $m_u/m_d=0.56$ \ref\baluni{V. Baluni,
\physrev{19}{1979}{2227}; R.J. Crewther, P. DiVecchia, G. Veneziano
and E. Witten; \pl{88}{1979}{123}; E \pl{91}{1980}{487} }. The
absence of strong CP violation violates `t Hooft's naturalness
condition that a parameter is only allowed to be very
small if setting it to zero increases the symmetry of the theory
\ref\thooft{G. 't
Hooft,  1979 Carge\`se Summer Insitute Lectures}. Simply setting
$\bar\theta$ to
zero requires fine tuning $\theta_{\rm QCD}$, the coefficient of
${\alpha_{\rm QCD}\over
8 \pi} G\tilde G$,  to cancel ${{\rm arg}\;{\rm det}\; m_q}$. Since CP
is broken by  the weak interactions, the fine tuning does not
increase the  symmetry of the standard
model.  This is the strong CP problem.

There are presently three possible solutions to the strong CP
problem, all of which involve imposing global symmetries on the low
energy world\foot{In addition there has been the proposal that
there is no strong CP problem---see  \ref\samuel{S.
Samuel, City College preprint CCNY-HEP-91-10 (1991)}; we do not see how
this solution could avoid leading to a light $\eta'$ meson. There
is also an argument that wormhole physics  may set  the effective
$\bar\theta$ to the CP conserving value of $\pi$ \ref\pieworms{
H.B. Nielsen
and M. Ninomiya, \prl{62}{1989}{1429}; K. Choi and R. Holman,{ \it
ibid.} 2575; \prl{64}{1990}{131};
J. Preskill, S.T. Trivedi and M.B. Wise, Phys. Lett. B323 (1989) 26
   }. }. One possible solution is that  the up quark is
massless.  In this  case there is an anomalous $U(1)$ symmetry at
the QCD scale,  rendering $\bar\theta$ unphysical. A massless up
quark is not necessarily in conflict with current algebra \ref\mrat{
S. Weinberg, Trans. N. Y. Acad. Sci 38 (1977) 185;  J. Gasser and H.
Leutwyler, Phys. Rep. 87 (1982) 77; H. Leutwyler,
\np{337}{1990}{108}}  since an effective up quark mass may be
mimicked by instanton    effects  \ref\mup{D.B. Kaplan and A.V.
Manohar, \prl{56}{1986}{2004}; K.  Choi, C.W. Kim  and W.K. Sze,
\prl{61}{1988}{794}; K. Choi, UCSD preprints UCSD-PTH-91-28 (1991),
UCSD-PTH-92-06 (1992)} or new low energy interactions \ref\annup{
A.E. Nelson, Phys. Lett.  254B  (1991)  282}.  Such theories require
that one impose a  chiral global symmetry (which is anomalous) to
eliminate the up quark's Yukawa coupling to the Higgs.

A second solution is that of Peccei and Quinn, who realized that an
anomalous
global $U(1)_{PQ}$ symmetry could be realized nonlinearly at low
energies,   still solving the strong CP problem while  allowing all
the quarks to get
masses from the Higgs \ref\pq{R.D. Peccei,
H. Quinn, \prl{38}{1977}{1440}, \physrev{D16}{1977}{1791}}.  This
nonlinear realization of $U(1)_{PQ}$ leads to
a  light pseudo Goldstone boson, the axion
\ref\axion{S. Weinberg, \prl{40}{1978}{223}; F. Wilczek, {\it ibid.},
279}. The axion couplings may be adjusted \ref\invaxion{
J.E. Kim, \prl{43}{1979}{103}; A. Vainshtein and V. Zakharov,
\np{166}{1980}; A.P. Zhitnitskii, Sov. J. Nucl. Phys. 31 (1980)
260;  M. Dine, W.  Fischler, and M. Srednicki, \pl{104}{1981}{199}}
to  avoid laboratory \ref\labc{F.P. Calaprice \etal,
\physrev{D20}{1979}{2708}; Y. Asano
\etal, \pl{107}{1981}{159}; S. Yamada, in Proc.1983 Lepton-photon
symposium (Aug 1983)
Cornell University}, astrophysical \ref\astroc{D.A. Dicus, E.W. Kolb,
V.L. Teplitz and  R.V.
Wagoner, \physrev{D22}{1980}{839}; M. Fukugita, S. Watamura and M.
Yoshimura,
\prl{48}{1982}{1522};  N. Iwamoto, \prl{53}{1984}{1198};
D. B. Kaplan, \np{260}{1985}{215};
M. Srednicki, \np{260}{1985}{689};
G.G.
Raffelt, Phys. Lett. 166B  (1986) 402;
A. Pantziris and K. Kang, \physrev{D33}{1986}{3509};
D.E. Morris, Phys. Rev. D34
(1986) 843;   G.G. Raffelt and D.S.P. Dearborn,
Phys. Rev. D36 (1987) 2211;  G. Raffelt and D. Seckel, Phys. Rev. Lett.
60 (1988) 1793;
M.S. Turner, {\it ibid.} 60 (1988) 1797; R. Mayle \etal \pl{203}{1988}{188};
W.C.  Haxton and K. Lee,  Phys. Rev. Lett. 66 ( 1991 )
2557}  and cosmological \ref\cosmoc{J.  Preskill, M.B. Wise and F.
Wilczek, \pl{120}{1983}{127}; L.F. Abbott, P. Sikivie, {\it
ibid.}, p. 133; M. Dine, W. Fischler, {\it ibid.}, p. 137}
constraints. Neither the axion nor the massless up quark solutions
are natural in the 't Hooft sense since in either case the global
$U(1)$ symmetry is only approximate---being anomalous---and
significant $U(1)$ violating instanton effects can occur that lead
to an unacceptably large $\bar \theta$ in the Peccei-Quinn case
\ref\ds{M. Dine and N. Seiberg, \np{273}{1986}{109};
J.M. Flynn and L. Randall, Nucl. Phys. B293 (1987) 731
}.

The third solution to the strong CP problem differs from the first
two in that instead of  an anomalous $U(1)$ symmetry, one imposes
an exact discrete symmetry---either CP or P---which is broken
spontaneously. The parameter $\bar\theta$ is then finite and
calculable, and in a special class of models   can be small enough
to be consistent with experiment\nref\spontcp{ A.E. Nelson,
\pl{136}{1984}{387}; S.M. Barr, \prl{53}{1984}{329}; S.M.
Barr, \physrev{D30}{1984}{1805},  Phys. Rev. D34 (1986) 1567}
\nref\spontp{K. Babu and R.N. Mohapatra, \physrev{D41}{1990}{1286};
S.M. Barr, D. Chang and G. Senjanovic,  \prl{67}{1991}{2765}}
\refs{\spontcp,\spontp}.

In this paper we will argue that only the third solution,
spontaneously broken CP (or P) may be viable  when
the effects of quantum gravity are considered. Quantum gravitational
effects arising from  wormholes
\ref\worm{S. Hawking, \pl{195}{1987}{337}; G.V. Lavrelashvili, V.
Rubakov, and P. Tinyakov,
JETP Lett. 46 (1987) 167; S. Giddings and A. Strominger,
\np{307}{1988}{854}; S. Coleman,
\np{310}{1988}{643}},   virtual black hole formation and evaporation
\ref\bh{S.W. Hawking,
\cmp{43}{1975}{199}}\ or nonperturbative effects in string theory
\ref\stringth{  M.B. Green, J.H. Schwarz and E. Witten,  {\it
Superstring Theory, Vol. 2}, Cambridge University Press, (1987),
and references therein}, may lead to violation of some or all
global symmetries in the effective theory  below the Planck mass
\ref\banks{T. Banks, Physicalia, 12 (1990) 19}. Presently  it is
unknown  how large the symmetry violating quantum gravitational
effects are or whether it is possible to escape the conclusion that
there are no exact global symmetries. In any case, both continuous
gauge symmetries and unbroken discrete subroups of gauge symmetries
(``discrete gauge symmetries'') are thought  to be preserved by
Planck scale physics, as they are violated neither by wormholes
nor  black holes \nref\discretehair{L. Krauss and F. Wilczek,
\prl{62}{1989}{1221}  }\nref\pk{J. Preskill and L.M. Krauss,
\np{341}{1990}{50}}\refs{\discretehair,\pk}. Approximate continuous
global symmetries may arise as accidental symmetries (symmetries
which are automatically respected by all gauge  invariant
renormalizable Lagrangians containing some set of fields), but are
otherwise expected to be violated by higher dimension operators
suppressed by inverse powers of the Planck mass $m_P$.

This leads one to contemplate a modified version of 't Hooft's
naturalness principle---which we will call ``strong
naturalness''---namely that all small parameters in the world must
be protected by {\it gauge}
symmetries.

Is the standard model consistent with strong naturalness?
Both lepton and baryon number in the standard model are examples of
accidental global symmetries, enforced by  $SU(3)\times SU(2)\times
U(1)$ invariance.  The lowest dimension allowed operators that
violate $B$ and $L$ are dimension six, which cause no problem if
suppressed by $m_P^2$. In the supersymmetric standard model there
are dimension four and five operators which cause problems such as
proton decay, but they can be eliminated by imposing discrete gauge
symmetries \ref\ir{L.E. Ibanez and G. Ross, \pl{260}{1991}{291},
\np{368}{1992}{3}}. The approximate global flavor symmetries that
arise in the  standard model due to the smallness of some of the
Yukawa couplings could also be protected from gravity by discrete
gauge symmetries  which are spontaneously broken below the Planck
mass.   The small value of the Higgs mass looks  unnatural in the
standard model, but in supersymmetric (SUSY) models, the soft SUSY
violating terms responsible for the Higgs potential could be protected
from gravitational
corrections by local supersymmetry at short distances (although there
remains the ``$\mu$-problem'', which must be solved by other means).
This leaves the cosmological constant (about which we have nothing new
to say) and $\bar \theta$ as the  remaining, curiously  small parameters
in the standard model.

At first sight, the strong CP problem does not appear to be
consistent with strong naturalness.  A vanishing up quark mass
cannot be protected by a discrete or continuous gauge symmetry, as
any  symmetry which protects only the up quark mass is anomalous in
the standard model---and the up quark Yukawa coupling would have to
be less than $ 10^{-14}$ for the EDMN to be unobserved.  The axion
solution cannot be protected by a gauge symmetry either, since the
$U(1)_{PQ}$ symmetry is also anomalous  (by construction) and cannot
be gauged. PQ violating operators arising from Planck mass
physics  can spoil the axion solution to the strong CP
problem\nref\sometime{H. Georgi, L. Hall, M. Wise, \np{192}{1981}{409}; G.
Lazarides, C. Panagiotakopoulos, Q. Shafi,
\prl{56}{1986}{432}}\nref\wyrm{J.E. Kim and K. Lee,
\prl{63}{1989}{20}}\nref\recent{
M. Kamionkowski and J. March-Russell, Princeton preprint
PUTP-92-1309 (1992); R. Holman \etal, Institute for Theoretical
Physics preprint  NSF-ITP-92-06 (1992);  S.M. Barr and D. Seckel,
Bartol preprint BA-92-11 (1992) } \refs{\sometime-\recent} (though also
see ref. \pieworms,
for a discussion of the effects of wormholes on $\bar\theta$ and the
axion potential).  If
Planck scale physics induces PQ-violating gauge invariant operators of
dimension d with coefficients $\CO(1/m_P)^{(d-4)}$,   then the axion
decay constant $f_a$ must be well below the Planck mass or there
will still be a strong CP problem even if axions exist. If $f_a$ is
above the astrophysical bound of $10^{10}$ GeV then the dimension
of the induced operators must be greater than ten. Models with an
accidental anomalous $U(1)$, respected by   all terms up to any
desired  dimension can be built \ref\acc{ J.A. Casas, G. Ross,
\pl{192}{1987}{119};  S.M.  Barr and D. Seckel in \recent, R.
Holman \etal\ in \recent; L. Randall, MIT preprint CTP-2050 }, but are
complicated and severely constrained.  In particular, such models
frequently have many colored particles and strong QCD interactions
at short distances, which means that small QCD instantons may
interfere with  the axion's ability to solve the strong CP problem
\ds.

Models with spontaneously broken CP can solve the strong CP problem
without requiring that any global symmetries other than CP itself
be imposed on the Lagrangian  \ref\noglob{S.M. Barr, in \spontcp,
P. H. Frampton and T.W. Kephart, Phys. Rev. Lett. 65 (1990) 820;  P.H.
Frampton and D. Ng,  Phys. Rev. D43 (1991) 3034,   also in the model
proposed by  L. Bento, G.C. Branco and P.A. Parada in   Phys.  Lett.
B267 (1991) 95, the only global symmetry imposed  besides CP is a
discrete symmetry  which could be a discrete gauge  symmetry }.
While this scenario satisfies 't Hooft's original naturalness
criterion, it fails under strong naturalness as CP could be
violated by   gravitational effects. For   example, black hole no
hair theorems imply that a black hole  which has swallowed a CP-odd
particle cannot be distinguished from one which swallows a CP-even
particle of the same mass.   However if CP were a discrete gauge
symmetry then  black holes  would carry discrete CP gauge hair
\refs{\discretehair,\pk}, and virtual black holes could not violate
CP. Discrete gauge symmetries are also not violated  by wormholes
\refs{\discretehair,\pk},  and should be preserved by all quantum gravitation
effects. In the next section we will show that CP can arise as a
discrete gauge symmetry, provided that there are more than four
spacetime dimensions. In \S3  we discuss the spontaneously broken
CP solution to the strong CP problem and restrictions on the scale
of spontaneous CP violation due to quantum gravity.  In \S4 we
examine one of the most amusing possible consequences  of CP as a
gauge symmetry: cosmic CP strings. Space  containing a CP string
has the curious property that a particle can be transported around
an uncontractible loop and transformed into its CP conjugate, and
that a four dimensional universe containing a   single string is
nonorientable, although the higher dimensional manifold {\it is}
orientable. In \S5 we comment briefly on the possibility of parity
as a discrete gauge symmetry.

\newsec{CP as a discrete gauge symmetry}

A discrete gauge symmetry arises when a continuous gauge symmetry G
is spontaneously broken to a subgroup H which has disconnected
components. A classic example \nref\classic{A.S. Schwarz,
\np{208}{1982}{141}}\nref\cole{ S. Coleman, 1975 Erice lectures,
published  in {\it Aspects of Symmetry}, Cambridge University
Press, (1985)}\nref\pres{ J. Preskill, 1985 Les Houches Lectures,
published in {\it Architecture of the Fundamental Interactions at
Short Distances}, North Holland, (1987)}\refs{\classic-\pres, \pk} is
the symmetry breaking pattern $SO(3)\rightarrow  O(2)$, which can be
realized by giving the traceless symmetric tensor representation of
$SO(3)$ a vev proportional to the matrix diag(1,1,-2). Such a vev
leaves a gauged $U(1)$ symmetry and a discrete charge conjugation
symmetry unbroken. Now charge conjugation is an element  of the
$SO(3)$ gauge group, which exchanges particles  with antiparticles
carrying the opposite $U(1)$ charge.

Can CP arise as such a discrete gauge symmetry? A major difference
between C and CP is that CP exchanges a left handed Weyl spinor
with its right handed complex conjugate. Since left and right handed
Weyl spinors are in different irreducible representations of the
Lorentz group, CP does not commute with Lorentz transformations. If
CP is to be an element of  a continuous local  symmetry which does
not commute with the four dimensional Lorentz group, then the Lorentz
group must be extended by introducing extra dimensions. CP may then be
embedded in this extended group.

Another apparent obstacle to imposing CP as a local symmetry is
that the usual CP transformation given by
\eqn\cptrans{
\eqalign{
\psi_L(x_\mu)&\to -\sigma_2\psi^*_L(x^\mu)\ ,\cr \phi(x_\mu)&\to \pm
\phi^*(x^\mu)\ ,\cr} \qquad
\eqalign{\psi_R(x_\mu)&\to \sigma_2\psi^*_R(x^\mu)\cr
A_\mu(x_\mu)&\to \pm A^\mu(x^\mu)}}
 reverses all
three    spatial dimensions as well as exchanging all fields with
their  complex conjugates. (In the above expression $\phi$, $\psi $,
and $A_\mu$   are (pseudo) scalar, fermion, and
 (pseudo) vector fields respectively, and
$\eta_{\mu\nu}=$diag$(+,-,-,-)$).  The reflection of the spatial
dimensions is not a symmetry of the Lagrangian, although it can be a
symmetry of the action; furthermore, it cannot be obtained by any local
transformation.

An alternative is to define a passive CP---denoted as \cp---
which does not involve
global spatial reflections.  It is defined as
\eqn\cphattrans{   \eqalign{x_\mu  &\to \bar{x}_\mu=x^\mu  \cr
\psi_L(x_\mu)&\to
-\sigma_2\psi^*_L(\bar{x}^\mu)= -\sigma_2\psi^*_L({x}_\mu) ,\cr
\phi(x_\mu)&\to\pm  \phi^*(\bar{x}^\mu)= \pm  \phi^*(x_\mu)\ ,\cr}
\qquad
\eqalign{&\cr \psi_R(x_\mu)&\to \sigma_2\psi^*_R(\bar{x}^\mu)
=\sigma_2\psi^*_R(x_\mu)\cr
A_\mu(x_\mu)&\to\pm A_\mu(\bar{x}^\mu)
=\pm A_\mu(x_\mu)\ ,}} where the first transformation is an
orientation changing general coordinate transformation. Note that
derivatives of fields transform as
 \eqn\derivtrans{
{\partial_\mu}\phi(x_\mu)\ {\mathop{\longrightarrow}\limits_{\widehat{{\rm
CP}}}}
\ \bar\partial_\mu  (\pm\phi^*(\bar{x}^\mu)
)= {\partial^\mu}  (\pm\phi^*(x_\mu) )\ .} Whereas under the usual CP
we have $\CL(x_\mu)\to\CL(x^\mu)$,   under  $\widehat{{\rm CP}}$
$\CL(x_\mu)\to\CL(x_\mu)$ for a CP invariant Lagrangian. However any
action which is invariant
under the usual definition of CP  is  invariant under  $\widehat{{\rm
CP}}$ and vice
versa. For example, the parity violating term  $\int {\rm d}^4x\
G\tilde{G}$ is eliminated from the  action since although this term
is   invariant under proper general coordinate
transformations, it is odd under coordinate
transformations with negative Jacobian, hence odd under \cp.
For the remainder of this paper we will drop the hat and simply refer to
the passive transformation \cphattrans\ as ``CP''.

We now contemplate  the possibility that the transformation
CP is an element of a continuous local symmetry $G$ of a
higher dimensional theory. The group $G  =G_L\times
G_g\times G_{YM}$, where   \hbox{$G_L$=spin($d-1$,1),} the
$d$-dimensional Lorentz group, $G_g=d$-dimensional general
coordinate transformations with positive Jacobian   and  $G_{YM}$
is the internal Yang-Mills group.  In order to have a chiral
fermion spectrum in four dimensions $G_{YM}$ must not be trivial
\ref\witten{M.F. Atiyah, F. Hirzebruch, in {\it Essays in Topology
and Related Subjects}, ed. A. Haefliger and R. Narasimhan
(Springer-Verlag, New York); E. Witten, \np{186}{1981}{412}; C.
Wetterich, \np{223}{1983}{109}; E. Witten, in {\it Shelter
Island II: Proceedings of the 1983 Shelter Island Conference on
Quantum Field Theory and the Fundamental Problems of Physics},
ed. R. Jackiw \etal, (MIT Press, Cambridge, Mass.)}.  We assume the
four dimensional CP transformation is given by the product of
$X_L(\in G_L)$, $X_g(\in G_g)$, $X_{YM}(\in G_{YM}) $ {\it i.e.}
\eqn\CPlocal{{\rm CP}=X_LX_gX_{YM}\ ,} and enumerate the
possibilities for $d$ and $G_{YM}$.

Note that alternative CP
transformations  are possible, which require  enlarging   the
particle spectrum. Any symmetry which reverses all gauge charges,
exchanges left and right handed fermions, and reverses the
orientation of space can be called CP. For instance the fermion
spectrum could be doubled by introducing mirror particles and CP
could exchange each fermion with its mirror. The restrictions of the
dimensionality of space and the gauge group which arise from
requiring that CP be embedded in a continous gauge symmetry will
depend on the definition of CP.     Since the observed
approximate CP exchanges all fields with their {\it own} complex
conjugates, we will   define CP by eq. \cphattrans.

 In order for the CP transformation to be an element of the local
symmetry group G of the underlying $d$-dimensional $( d > 4)$ theory
 we need  all four dimensional fields to be in the same
irreducible representation  of G    as their complex
conjugates. If both
$G_L$ and $G_{YM}$ admit an inner automorphism (an automorphism
which  is an element of the group) which exchanges each element of
a representation with its complex conjugate, then we identify
these automorphisms as $X_L$ and $X_{YM}$ respectively.  This is only
possible if the Yang-Mills group $G_{YM}$  is one of the Lie  groups $E_8$,
$E_7$,
$SO(2n+1)$, $SO(4n)$, $Sp(2n)$, $G_2$ or $F_4$ (or a product of these
groups). Only these groups admit an inner automorphism which changes
the sign of the entire Cartan subalgebra \ref\slansky{R. Slansky,
Physics Reports C79 (1981)}.
For the Lorentz group,
the  irreducible spinor representations  must
be Majorana. For an even number of spacetime dimensions, we
require that the Majorana condition commutes with the Weyl
condition, {\it i.e.} we need Majorana-Weyl spinors. (If the
Majorana and Weyl
conditions do  not commute  then  a Majorana
spinor  is the direct sum of a complex Weyl spinor and its
conjugate, and   a Lorentz transformation cannot exchange the Weyl
spinors with their complex conjugates.) Majorana spinors are allowed
for odd dimensional spacetimes of dimension $d=8 k+1, 8k+3$ and
Majorana-Weyl spinors are only possible in $8k+2$ dimensions
\ref\majspin{G. Chapline and R. Slansky, \np{209}{1982}{209}; C.
Wetterich, \np{211}{1982}{177}},  and for these dimensions   the
Lorentz group contains an inner automorphism which complex
conjugates all representations. For  $d=8 k+1,8k+2,$ and $ 8k+3$,
and for the Yang-Mills groups specified above, we can define four
dimensional CP to be the product of an    inner automorphism of
$G_{YM}$, a $G_g$ transformation $X_g$ which
reverses the orientations of both 4-D Minkowski space and the
compactified manifold, and the $G_L$ Lorentz transformation
\eqn\lt{X_L= \left(\vbox{ \halign{
\hfil $#$ \hfil&\vrule#&\hfil $#$ \hfil\cr
\lorentz && \mzer \cr \multispan3\hrulefill\cr
\mzer && \mK \cr}\vskip-.6truein}\right) }
where $K_L$ is a $(d-4)$ dimensional real matrix satisfying $K_L
K_L^T=1$ and det$(K_L)= -1$.   $X_g$ must be chosen to include a
compactified coordinate transformation
$\theta_i\rightarrow-\theta_i$, where the vector fields
$\partial/\partial\theta_i$ generate the Cartan subalgebra of the
continuous isometry group of the compactified dimensions. Then the
transformation CP=$X_LX_gX_{YM}$ exchanges all gauge charges of
four dimensional fields with the gauge charges of their
antiparticles, including gauge interactions arising from the
 isometries of the compactified space.  Both $X_L$ and
$X_g$ are orientation preserving on the  higher dimensional space
and   are elements of the   local symmetry group of the full
theory.

An   interesting case where CP can arise as a discrete
local symmetry occurs in superstring theory
\stringth.
The popular 10 dimensional superstring theories with
gauge  group $SO(32)$ or $E_8\times E_8$    satisfy the
conditions given above for  four dimensional CP to be a gauge
transformation. These
are also examples of higher dimensional theories which can have an
acceptable low energy particle spectrum and  gauge group, for some
compactifications. Furthermore, it is possible for CP  to
remain unbroken after compactification in realistic models. For
instance, as was noted by Strominger and Witten
\ref\stringcp{A. Strominger and E. Witten,
\cmp{101}{1985}{341}; C.S. Lim, \pl{256}{1991}{233}}
 when the $E_8\times E_8$ superstring has six dimensions compactified
on an internal manifold of $SU(3)$ holonomy, with the spin
connection  embedded in the gauge connection,
the existence
of a discrete orientation reversing isometry of the internal manifold
guarantees CP invariance of the four dimensional theory.
 Furthermore  for some compactifications the four
dimensional effective theory may also contain extra gauge symmetries
({\it e.g.}  additional $U(1)$'s or a four dimensional grand unified
$E_6$ gauge symmetry) and fermion fields (additional quarks which
are vector-like under the low energy $SU(3)\times SU(2)\times U(1)$
gauge symmetry with  the quantum numbers of a right handed down
quark  plus its mirror). These additional fields and
gauge symmetries could play the role of the new particles and
symmetries introduced in \nref\superspont{ For
a  discussion of spontaneously broken CP in supersymmetric models see
S.M. Barr, A. Masiero, Phys. Rev. D38 (1988)
366}refs.~\refs{\spontcp,\noglob,\superspont}
to constrain the form of the fermion mass matrices, as needed to
solve the strong CP problem in spontaneously broken CP models\foot{
It is commonly thought that string theories can solve the strong CP
problem by means of the axion mechanism \ref\stax{E. Witten,
\pl{149}{1984}{351}}, but it is far from clear that
the mechanism survives nonperturbative contributions to the axion
potential from hidden sector gauge interactions \ds\ and string world
sheet instantons \ref\stringy{X.G. Wen and E. Witten,
\pl{166}{1986}{397}; M. Dine, N. Seiberg, X.G. Wen and E. Witten,
\np{287}{1986}{769}}.}.

Four dimensional C, P, and  CP invariance arising from higher
dimensional theories has been discussed  before
\nref\kkcp{W.  Thirring, Acta Phys. Austr. Suppl 9 (1972) 256;   C.
Wetterich, \np{234}{1984}{413};
M.B. Gavela, R.I. Nepomechie, Class. Quant. Grav. 1 (1984)
L21}
\refs{\stringth,\kkcp,\stringcp} for
the purpose of ruling out some higher dimensional theories.
However this  earlier work on CP invariance in higher dimensional
theories has been motivated by a goal  orthogonal to ours, namely
it has been assumed that CP should {\it not} be a symmetry   of the
effective four dimensional theory.  What we are proposing is that
CP remains an exact discrete gauge symmetry below the
compactification scale and is spontaneously broken at much longer
distances, accounting for the
 violation observed in the weak interactions. The CP
violating vacuum  of the full theory can be considered to be a
perturbation from the CP conserving configuration. If the deformation is
small enough and can be described by four dimensional scalar fields,
then we can study the CP conserving effective four dimensional
theory which arises from compactification to the CP conserving
pseudo-vacuum. This effective theory includes CP odd scalar fields
whose vevs are non zero in the ground state.   Planck scale physics
will not induce CP violating operators in the effective theory,
although in general we do expect non renormalizable CP conserving
operators, suppressed by powers of the compactification scale, which
we take to be near the Planck mass. We discuss the possible effects
of such operators in the next section.

\newsec{Spontaneously broken CP and the strong CP problem}

If CP is spontaneously broken then the strong CP parameter
$\bar\theta$ is   calculable, but not necessarily small.
\nref\thetacalc{A.E. Nelson, \pl{143}{1984}{165}}  A class of
models in which $\bar\theta$ does come  out to be naturally small, in
the sense of 't Hooft,  is described in
refs.~\refs{\spontcp,\superspont,\thetacalc}.   In these papers
only renormalizable Lagrangians are  considered, and the issue of
strong naturalness is not discussed, since CP is assumed to be an
exact  global symmetry. However, as we have just shown, CP could be
a discrete gauge symmetry which is left unbroken at the
compactification scale, and then quantum gravitational effects
could not induce CP violating operators in the effective four
dimensional theory. However there could be nonrenormalizable terms
arising from Planck scale physics and suppressed by powers of the
Planck mass, which will lead to additional CP violating effects
below the scale of spontaneous CP violation. Here we show that
inclusion of such nonrenormalizable operators in the effective
Lagrangian may spoil the spontaneously broken CP solution to the
strong CP problem. We will argue that this solution should survive
Planck scale effects if  the scale of spontaneous CP violation is
below  $10^9$ GeV.

First let us review a class of models utilizing the spontaneously
broken CP solution to the strong CP problem.  Models which satisfy
the following criteria can give the standard model with sufficiently
small strong CP violation as a low energy effective theory, when the
effects of Planck scale physics are ignored.

\item{1.} The   ordinary Higgs doublet, $H$,   acquires an
$SU(2)\times U(1)$ breaking vev  whose phase  may be chosen to be
zero (In multi Higgs doublet models   if the relative phases of the
Higgs doublets are nonzero the neutron electric dipole moment may
come out too large   \ref\multidoublet{R. Akhoury and I.I. Bigi,
\np{234}{1984}{459}; D. Chang, K. Choi, W.Y.
Keung, \physrev{D44}{1991}{2196}}). The Higgs couples with real
Yukawa couplings to $n_f$ families of  type ``$F$'' quarks. The
$F$ quarks are in ordinary chiral representations of
$SU(3)\times SU(2) \times U(1)$.

\item{2.} CP is broken spontaneously by the
vevs of some complex scalar fields $\phi_i$ which are singlets under
$SU(3)\times SU(2) \times U(1)$. The ordinary gauge  symmetries
prevent any tree level Yukawa couplings of type  $F$-$F$-$\phi_i$.

\item{3.} There is another set of quarks called $C$ and $\bar
C$, distinguished from the $F$--type quarks by new
symmetries.  In
ref.~\spontcp\ additional global symmetries were    used, but
these may easily be gauged.   The $C$-type quarks have ordinary
$SU(3)\times SU(2) \times U(1)$ quantum numbers, and the $\bar C$'s
have mirror quantum numbers, {\it i.e.}
$SU(3)\times SU(2) \times U(1)$ allows a mass term connecting  $C$
and $\bar C$. However this mass term must be real, {\it i.e.} any
coupling of form $\bar C$-$ C$-$\phi_i$ is forbidden by the new
symmetries.

\item{4.}Yukawa couplings of type $\bar C$-$F$-$\phi_i$
are allowed, which is how CP violation is communicated to the ordinary
quarks.

\item{5.}No Yukawa couplings of type $C$-$C$-$H$ are allowed by the
new symmetries.

 In this
class of models  at tree level the quark mass matrix has the form
\eqn\barrmatrix{\bordermatrix{ &\hfill {F_R} \hfill & \hfill
{C_R} \hfill & \hfill {\bar C_R}\hfill\cr
F_L&{\rm real}& 0 & {\rm complex}\cr
 &&&\cr C_L&0& 0 & {\rm real}\cr
 &&&\cr \bar C_L&{\rm complex}&{\rm real} & {\rm complex }\cr}\;\;
,}  which   has real determinant.

The masses involving the $C$ and
$\bar C$ quarks do not break $SU(3)\times SU(2) \times U(1)$ and may be
much  larger than the weak scale. There will always be at least $n_f$
families of light quarks with weak scale masses, which may be
complex linear combinations of $F$ and $C$-type quarks. All   $\bar
C$ quarks may get heavier than the weak scale by combining with the
orthogonal   combinations of $F$ and $C$ fields. At low
energies the light quark mass matrix will be complex, and ordinary
weak CP violation can occur via the usual unremovable phase in the
Kobayashi-Maskawa weak mixing matrix. If the CP-violating $\phi_i$ vevs
are big and  the  $\bar C$ quarks fairly heavy (at least $\CO$(TeV)
 ) then the low energy effective theory can be the standard model
and  flavor changing neutral currents are suppressed.

 For a discussion of radiative corrections to $\bar\theta$ in these
models see ref.~\thetacalc; for the radiative corrections in the
supersymmetric case see ref.~\superspont. These radiative corrections
can easily be as small as $\CO(10^{-11})$.

Note that the determinant of \barrmatrix\ is real only if the mass
term connecting $C$ and $\bar C$ is real, which may be guaranteed by
a gauge symmetry allowing $\bar{C}$-$C$ terms but forbidding
$\bar{C}$-$C$-$\phi_i$ Yukawa couplings.
In  general, however, such a symmetry does not forbid
$\bar{C}$-$C$-$\phi_i$-$\phi_j^*$
couplings. These couplings are nonrenormalizable, and so suppressed by
$m_P$, but can give the $\bar{C}$-$C$ mass term a phase proportional
to $\vev{\phi_i }/m_P$. Thus we expect the effective tree level
$\bar\theta$ to receive a contribution of order
\eqn\barrtheta{\bar\theta=\CO\left({\vev{\phi_i }\over m_P}\right),}
which will be smaller than  $10^{-10}$ if the scale of spontaneous CP
violation is below $10^9$ GeV.

\newsec{CP strings and walls}

Imagine returning from a long celestial journey, only to find
your loved ones to be made of antimatter, with their hearts
on the wrong side. Apparently you travelled around a cosmic ``CP
string''. These CP strings could exist as
defects if CP were an unbroken discrete gauge symmetry.
Fortunately,   CP strings cannot survive below the scale of
spontaneous CP breaking, but they may   play an important role in
the early universe by eliminating cosmologically undesirable  CP
domain walls \ref\domwal{Ya. B. Zel'dovich, I.Yu. Kobzarev, L.B.
Okun, JETP 40 (1975) 1}.

Topologically stable strings in four spacetime dimensions are
possible whenever a  simply connected  internal gauge group G breaks
to a group H containing disconnected components
(stable strings are possible whenever the manifold
G/H is not simply connected \cole). An amusing example is the
``Alice string'' \refs{\pk,\classic,\pres} which arises when charge
conjugation is an element of a spontaneously broken gauge group.
In the presence of an Alice string, electric charge is double
valued. Furthermore the relative charge between two particles can be
changed by sending one of them around the string.

Here we consider the  case where the action is invariant under a
symmetry group G, which includes general coordinate transformations
in $d>4$ dimensions as well as internal symmetries. We assume the
vacuum  configuration spontaneously breaks G to
a  group H which includes   CP as a discrete element.
 Since CP
is an unbroken element of a spontaneously broken continuous symmetry
then stable CP strings should exist \ref\prestwo{
   ``CP strings'' and their cosmological properties,  were
described by John Preskill in a talk at the ``Topical Conference on
Cosmological Phase Transitions'' given at the ITP, Santa Barbara
(April 3, 1992). The  strings discussed by Preskill differ from ours in
that they arise  from gauge
symmetry breaking in ordinary  four dimensional models with
orientable four dimensional spacetime. }.

A CP string is a  defect
in the spacetime manifold, as well as in the internal gauge group
symmetry breaking order parameter.
An object parallel transported around such a string could experience, for
example,  a    $180^o$ rotation in a plane containing one
compactified and one ordinary dimension,  while having all gauge charges
are conjugated as with the Alice string.  When two explorers with right-handed
coordinate systems start from the same point in space,  pass around opposite
sides of a CP string,  and reencounter each other on the other side,
  each one will think that the other person has been CP conjugated,
while claiming that they remain unchanged themselves.  To bolster her
claim that she remains unchanged, explorer B points to the right-handed
reference
frame she has brought with her; however explorer A sees her hold up a
left-handed reference frame constructed out of antimatter, and isn't
convinced. In fact, to claim that one has changed and the other has not
is not a gauge invariant statement.
In order to avoid potentially dangerous physical conflict,
the two explorers agree to set up an imaginary surface ending on the
string, with the convention that anyone crossing this surface redefines
``left-handed'' to mean ``right-handed'' and ``matter'' to mean
``antimatter''. This is entirely analogous to the international dateline, where
we reset our watches by 24 hours,  and like the dateline, its position
is a matter of convention.  Now when the explorers meet, they can agree
on the definitions of  matter  and antimatter,  left-handed and
 right-handed.  We will refer to
this surface as the ``CP dateline'' \foot{The CP dateline is entirely
analogous to the
Preskill-Krauss construction for Alice strings \pk.}.
Note that in the presence of such a string, the full manifold---including
the compactified dimensions---is orientable, while
the four-dimensional spacetime is not.  Such strings have many peculiar
properties \ref\cpstr{A.G. Cohen, D.B. Kaplan, A.V. Manohar and A.E.
Nelson, in preparation.}.

 The CP string should be very heavy, having
mass/length comparable to the compactification scale squared.
Inside the string, at least some of the higher dimensional
symmetries and extended gauge symmetry are realized. It is not
clear whether such strings   ever get formed in the early universe,
since the compactification of the extra dimensions is not
necessarily a phase transition, but if they are formed   their
 cosmological behavior could be similar to other gauge strings which
have been proposed to seed galaxy formation and large scale
structure    \ref\cosmicstring{A. Vilenkin,
\physrev{D23}{1981}{852}; A. Albrecht, N. Turok,
\prl{54}{1985}{1868}; D.B. Bennett, F.R. Bouchet,
\prl{60}{1988}{257}}, and, if they survive until recent epochs they
must be characterized by a string tension less than $10^{-5}m_P^2$
so that they do not contribute large inhomogeneities to the
microwave background \ref\micro{A. Stebbins, Ap. J. 327 (1988) 584;
D. Bennett, A. Stebbins, Nature 355 (1988) 410}. In many models of
compactification it is difficult to see how  strings could be
produced  without also producing heavy  magnetic monopoles, which
are troublesome cosmologically as they tend to contribute too much
mass density to the universe \ref\mono{ Ya. B. Zel'dovich, M. Yu.
Khlopov,  Phys. Lett. 79B (1978) 239; J.P. Preskill, Phys. Rev.
Lett. 43 ( 1979) 1365  }, however the monopoles may be eliminated
by either inflation (which also eliminates the strings)
\ref\inflation{ A.H. Guth, \physrev{D23}{1981}{347}; A.D. Linde,
Rep. Prog. Phys. 42 (1979) 389; A.D. Linde, \pl{108}{1982}{389}; A.
Albrecht, P.J. Steinhardt, \prl{48}{1982}{1220}  } or the
Langacker-Pi  mechanism \ref\langpi{P. Langacker, S.-Y. Pi, Phys.
Rev. Lett. 45 (1980 ) 1}.

The era of CP strings ends when some complex scalars acquire CP
violating vevs, and the strings become attached to domain walls. One
must distinguish between the effects of the wall, and those of the CP
dateline surface; they can be chosen to be coincident by convention, but it
is not necessary. We will consider them {\it not} to be coincident, for
pedagogical reasons. On traversing the domain wall but not the dateline,
the CP violating phase in the scalar vev changes sign, and therefore so does
the
 the CP violating phase in the low energy Hamiltonian.  For
example, consider an experiment involving kaons in the vicinity of the
domain wall.  When a $K_{\rm L}$ meson, with wavefunction
\ref\pdb{Review of Particle
Properties, \pl{239}{1990}{VII.90}}
\eqn\klong{\ket {K_{\rm L}}= { (1+\epsilon)\ket {K^0} - (1-\epsilon) \ket {\bar
K^0} \over \sqrt{2(1+\vert\epsilon\vert^2)}}}
 is transported rapidly (\ie\ nonadiabatically) across the domain wall, its
wavefunction is the same on the far side, but it
is no longer  a mass eigenstate, since on the far side of the wall
there is a different Hamiltonian: $H(-\epsilon)$ instead of
$H(\epsilon)$.
What was a $K_{\rm L}$ is now  a superposition of
 the $\pr{K_{\rm L}}$ and $\pr{K_{\rm S}}$ mesons, which are the mass
eigenstates of $ H( -\epsilon)$.  This is a physical effect:
the $\pr{K_{\rm S}}$ component decays away quickly and can be observed.
The remaining $\pr{K_{\rm L}}$ component will eventually decay as well,
but one will notice that its CP violating leptonic decay mode  favors
electrons over the usual positrons. The experimenter who carried
the meson through the domain wall will still be made of
electrons and nucleons.
If a second experimenter carries a second $K_{\rm L}$ meson from the
same initial point to the same final point, choosing a trajectory that
does not pass through the domain wall, but instead circumnavigates the
string at the wall boundary, she will pass through the CP
dateline.  At the dateline $\epsilon\to -\epsilon$ both in the
hamiltonian and the meson wavefunction, by convention. The kaon will
remain a mass eigenstate, and will also decay
preferentially into electrons (with greater amplitude than the first
meson, since there is no  $\pr{K_{\rm S}}$ component), but this
experimenter is now made of positrons and antinucleons.  That the two
mesons behave differently is a gauge invariant fact: one had a component
decay with the $K_{\rm S}$ lifetime while the other didn't.  This is
acceptable because the domain wall, unlike the CP dateline, is a
physical barrier which one meson traversed (nonadiabatically) and the other did
not.

The cosmological properties of walls
ending on strings have already been discussed  \nref\kls{T.W.B.
Kibble, G. Lazarides, Q. Shafi, \pl{113}{1982}{237},
\physrev{D26}{1982}{435}}\nref\vilev{A. Vilenkin, A.E. Everett,
\prl{48}{1982}{1867}}\nref\vilenkin{A. Vilenkin, Phys. Rept.
121 (1985) 263}\nref\ptww{ J. Preskill, S. P. Trivedi, F.
Wilczek, M. B. Wise,  Nucl. Phys. B363 (1991) 207
}\refs{\prestwo,\kls-\ptww}.  Neither the strings nor the domain
walls are topologically stable below the CP breaking scale.   A
single domain wall is metastable, since, although holes  bounded by
strings can appear in the wall, the probability of a hole
appearing which is large enough to grow is exponentially suppressed
by  \eqn\expon{\exp(-\CO( \mu^3/\sigma^2))\ ,} where $\mu$ is the
string tension and $\sigma$ is the wall tension. However if strings
left over from a compactification transition are present at the
scale of spontaneous CP violation, the  resulting network of walls
and strings   soon vanishes, unless a period of inflation occurs
between the two phase transitions.

\newsec{Parity as a gauge symmetry}

Spontaneously or softly broken parity has also been proposed as a
solution to the strong CP problem \spontp.  The usual four
dimensional parity transformation can be embedded in the local
general   coordinate invariance and Lorentz symmetries, provided
there are five or more dimensions. However models with unbroken
ordinary four dimensional parity are not phenomenologically viable
since the observed fermions are in a chiral representation of the
gauge group. This problem cannot be alleviated via spontaneous
parity breaking. However the product of space-time parity and an
internal symmetry transformation may also be called parity, as in
left-right symmetric models \ref\leftright{J.C. Pati, A. Salam,
\physrev{D10}{1974}{275}; R.N. Mohapatra, J.C. Pati,
\physrev{D11}{1975}{566}; G. Senjanovic, R.N. Mohapatra,
\physrev{D12}{1975}{1502}} where the parity transformation also
exchanges two inequivalent $SU(2)$ gauge groups. In these theories
parity does not commute with internal gauge transformations and
fermions can be in chiral representations of the gauge group.
  Such a four dimensional
theory could conceivably arise from the compactification of a higher
dimensions.   Four dimensional parity
is then a product of an inner automorphism of the
Yang-Mills group, a  local Lorentz transformation, and a
 general coordinate transformation of the higher dimensional
theory.  Obtaining
fermions   in chiral representations of the four dimensional group
places severe restrictions on the number of extra dimensions  and
the higher dimensional theory \refs{\witten,\kkcp}. An interesting
feature of this sort of parity transformation is that for some
choices of the spectrum and gauge groups there is no
phenomenological requirement for parity to be broken
\ref\unbrokenp{R. Foot, H. Lew, R.R. Volkas, University of Melbourne
preprints UM-P-91/54 (1991), UM-P-91/82 (1991)}. In the unbroken
parity models parity is a product of the usual parity and the
exchange of the usual $SU(3)\times  SU(2)\times U(1)$ gauge group
with a ``shadow'' $SU(3)\times  SU(2)\times U(1)$.  If the parity
transformation of such models is an element of a continuous gauge
symmetry then it is possible that parity strings are topologically
stable and could exist in our current universe. Circumventing
such a parity string would be quite disconcerting, as not only would
you find your loved ones with their hearts on the wrong side, but
they would  be    invisible, since they would now   be
made of matter which doesn't carry the same gauge
interactions as the matter you are made of\foot{J. Preskill, in
\pres,  remarked on a similar possibility in connection with
unbroken charge conjugation.}.

 \newsec{Summary}

  Planck mass physics effects such as wormholes may violate global
symmetries, including the Peccei-Quinn symmetry usually invoked to
solve the strong CP problem.   However there exists a solution to
the strong CP problem  which does not involve any global
symmetries--namely that CP  (or P)  could be a discrete gauge
symmetry which is spontaneously broken below well the Planck mass.
If the spontaneous CP violation scale is lower than $10^9$ GeV, then
Planck scale physics should have little effect on this solution.
With such a low scale  for spontaneous CP   violation one must
worry about several cosmological problems. In general, the
spontaneous breaking of a discrete symmetry produces cosmologically
undesirable domain walls. Domain  walls produced at the spontaneous
CP or P violation scale will not be topologically stable, since the
discrete symmetries are embedded in a continuous symmetry, but can
be metastable. There are at least two possibilities for avoiding a
cosmological  domain wall problem; either inflation occurs and the
reheating temperature after  inflation is below the scale of
spontaneous CP violation, or the domain walls  can end on cosmic CP
strings left over from an earlier stage of symmetry breaking  and
the entire network of walls and strings will  disappear. Baryogenesis
must  take place during or after spontaneous CP violation; there
are  many possibilities for low energy baryogenesis
\ref\dolgov{For a recent  review see A.D. Dolgov, Kyoto preprint
YITP-K-940 (1991)  }.

 \bigbreak\bigskip\bigskip\centerline{\bf
Acknowledgments}\nobreak
We would like to thank Jeff Harvey for interesting us in the
question of how CP arises from higher dimensional theories.  We
also thank Tom Banks, Andy Cohen, Aneesh Manohar, Michael Peskin and
John Preskill for conversations.   This work is supported in part by
DOE contract \#DE-FG03-90ER40546. A.N. and D.K. are supported in
part by grants from the Alfred P. Sloan Foundation. D.K. is also
supported in part by NSF contract PHY-9057135, and A.N. is supported
in part by a fellowship from the Texas National Research Laboratory
Commission.

\listrefs

\end\bye

\appendix{A}{}
In this appendix we describe in more detail how CP can arise as an
element of a higher dimensional symmetry group. As usual we start by
asuming that the  spacetime manifold is $M_4\times K$
where $M_4$ is flat four dimensional Minkowski space and $K$ is a
compact manifold.    In conventional Kaluza-Klein theories
  a subgroup of the general coordinate
transformations is unbroken if the  configuration is
form invariant  under the corresponding transformation, {\it i.e.}
if the metric tensor $g_{ab}(x)$ transforms as
\eqn\mettrans{g_{ab}(x)\rightarrow
\bar{g}_{ab}(\bar{x})=g_{ab}(\bar{x}).}  Here we want
the metric tensor to be form invariant  under the transformation CP
which is a product of a general coordinate transformation $X_g$, a
local Lorentz transformation $X_L$, and an internal Yang-Mills
transformation $X_{YM}$. The metric tensor  is only affected by
$X_g$,   under which
 the $M_4$  coordinates transform as $x_\mu\rightarrow x^\mu$, and
the internal coordinates $y$ transform as $y\rightarrow\bar{y}$.
Clearly the transformation $y\rightarrow\bar{y}$ must be an
orientation changing isometry of the internal manifold. Since we
are not only considering the case of pure Kaluza-Klein
theories, but are in general  including fermion and other
matter fields, and an internal Yang-Mills group, for  CP to be
an unbroken symmetry, we need not only the form invariance of
the metric, but all     fields must be form invariant
under the combined transformation \CPlocal\ in the  vacuum
configuration. The vielbein $e^a_m(y)$ and the  gauge field
configurations $A_m(y)$ on the compactified space
must transform as
\eqn\vieltrans{X_gX_L:\  e^a_m({y})\rightarrow
(K_L)^a_b{\partial{y}^n\over\partial\bar{y}^m} e^b_n(y)=
e^a_m(\bar{y})} \eqn\gaugetrans{X_gX_L:\ A_m({y} )\rightarrow
 {\partial{y}^n\over\partial\bar{y}^m}A_n(y) =-A^*_m(\bar{y})
}under the product of the general coordinate and local Lorentz
transformations. Then since   under the   Yang-Mills transformation
$X_{YM}$    $A_m(y)\rightarrow -A^*_m(y)$, under \CPlocal\ the
  gauge field configuration transforms as
\eqn\totalgaugetrans{ X_gX_LX_{YM}:\  A_m({y})\rightarrow
 -{\partial{y}^n\over\partial\bar{y}^m}A^*_n(y)= A_m(\bar{y})\ .}

For suitable choices of the number of dimensions and the internal
Yang-Mills group, we can find a transformation of the form
\CPlocal\ which effects the CP transformation in the four
dimensional world. In order for this transformation to be called
CP, it must exchange the gauge charges of particles and
antiparticles, for gauge charges which arise either from  the
unbroken subgroup of the $d$-dimensional Yang-Mills group or those
arising in Kaluza-Klein fashion from the isometry group of the
compactified space. We can choose CP  to complex conjugate    the
gauge representations, including those gauge symmetries arising
from isometry groups. The internal space isometry includes both a
general coordinate transformation and the local Lorentz
transformation, such that the vielbein is invariant under the
combined transformations. The transformation \lt\ includes a
transformation of the internal space Lorentz group spin$(d-4)$
which changes the sign of the entire cartan subalgebra, hence
exchanges all fermion representations with their complex
conjugates.  The transformation $X_g$ is chosen to include a
compactified coordinate transformation
$\theta_i\rightarrow-\theta_i$, where the vector fields
$\partial/\partial\theta_i$ generate the Cartan subalgebra of the
isometry group of the compactified dimensions. Then the
transformation CP=$X_LX_gX_{YM}$ exchanges all gauge charges of
four dimensional fields with the gauge charges of their
antiparticles.

The CP transformation on four dimensional spinor representations
of the local Lorentz group  is  also easily reproduced. In $d= 8
k+1,8 k+2$ or $8k+3$ dimensions a Majorana or Majorana-Weyl spinor
$\Psi(x,y)$ can be expanded as
\eqn\psiexp{\Psi(x,y)=\sum_i(\psi_{iL}(x)\otimes\eta_{i }(y)+
\psi_{iR}^c(x)\otimes\eta_{i }^c(y))\ . } Here the $\psi_{iL}$
are four dimensional  Weyl spinors and $\psi_{iR}^c$ are their
charge conjugates.  The $\eta_{i }$ are   $(d-4)$ dimensional
spinors  which form a complete set of spinor harmonics $\eta_i(y)$
in the internal space, and  and $\eta_{i }^c$ are their charge
conjugate spinors. (For the odd-dimensional cases $d=8k+1,8k+3$,
the $\eta_{i }$ are equivalent to $\eta_{i }^c$; in the case
$d=8k+2$ $\eta_{i }$ and $\eta_{i }^c$ have opposite helicity and
are equivalent to complex conjugates of each other.) The
$d$-dimensional gamma matrices $\Gamma_a\ (a=0,1,. . .,d-1)$ take
the form \eqn\gammaform{ \Gamma_\mu=\gamma_\mu\otimes I,\
\Gamma_i=\gamma_5\otimes\tilde\gamma_i\ ,}where the $\gamma_\mu$
are the usual four dimensional gamma matrices, $\gamma_5$ is the
usual four dimensional chirality matrix, and the $\tilde\gamma_i$
are  gamma matrices for the spin$(d-4)$ symmetry of the
compactified dimensions. In the spinor representation the Lorentz
transformation $X_L$ of eq.~\lt\
is\eqn\spinorspace{\Gamma(X_L)=\gamma_0\otimes\tilde\gamma\ ,}where
$\tilde\gamma$ is a product of the internal space gamma matrices
which flips the internal space helicity for the case $d=8k+2$. Thus
$\Gamma(X_L)$ exchanges $\psi_{iL}$ with  $\gamma_0 \psi_{iR}^c$,
as desired for four dimensional spinors.

Here we have argued that for certain higher dimensional theories
one can   find an  element of the local symmetries which can
be identified with the four dimensional CP transformation after
spontaneous compactification. Howver this transformation gives not
only the CP transformation of the four dimensional fields but also
a transformation of generic internal space harmonics, {\it eg.}
harmonics which appear as coefficients of bosonic four dimensional
fields. In the four dimensional effective theory, the
transformation of the internal space harmonics will give rise to a
complex conjugation of the coupling constants in the effective
Lagrangian.

When the phases in the coupling constants can be removed by
redefinitions of four dimensional fields, than CP is not
spontaneously broken and is an exact discrete gauge symmetry. We
are interested in the case where CP is preserved during the
compactification and then spontaneously broken by the vevs of four
dimensional scalar fields, in order to solve the strong CP
problem. The CP violating vacuum  of the full theory can be
considered to be a deformation of a CP conserving pseudo-vacuum. If
the deformation is small enough and can be described by four
dimensional scalar fields then we can study the CP conserving
effective four dimensional theory which arises from
compactification to the CP conserving pseudo-vacuum. This
effective theory includes CP odd scalar fields whose vevs are non
zero in the ground state.   Planck scale physics will not induce CP
violating operators, although in general we do expect non
renormalizable CP conserving operators in the effective theory.

An interesting example of how CP can arise as an
unbroken discrete gauge symmetry in four dimensions is the
$E_8\times E_8$ ten dimensional superstring spontaneously
compactified on an internal manifold of $SU(3) $ holonomy with the
spin connection $\omega_m$ embedded in the gauge group, {\it
i.e.}\eqn\spincon{\omega_m=A_m\ (\in SU(3))\ .} This case was
studied  in ref. \stringcp. The mere existence of a discrete
orientation reversing isometry of the Calabi-Yau manifold guarantees
that  eq. \gaugetrans\ is   satisfied since given equations
\vieltrans\ and \spincon\ we must also have
\eqn\calabiyau{\eqalign{ \omega_m(\bar{y})=&
{\partial{y}^n\over\partial\bar{y}^m}(K_L)\omega_m(y)(K_L)^{-1}\cr
=&-{\partial{y}^n\over\partial\bar{y}^m}\omega_m^*(y)\cr
\rightarrow A_m(\bar{y})=&
-{\partial{y}^n\over\partial\bar{y}^m}A_m^*(y) \ . \cr } }
 Note that in this model, for some compactifications, the four
dimensional effective theory may also contain extra gauge symmetries
(either additional $U(1)$'s or a four dimensional grand unified
$E_6$ gauge symmetry) and fermion fields (additional quarks which
are vector-like under the low energy $SU(3)\times SU(2)\times U(1)$
gauge symmetry with  the quantum numbers of a right handed down
quark  plus its mirror). These additional fields and
gauge symmetries could play the role of the new particles and
symmetries introduced in refs.~\refs{\spontcp,\noglob,\superspont}
to solve the strong CP problem.

\listrefs

\end\bye